# Cataract Vision Mimicked By Means Of Protein Denaturation In Egg Albumen

B. Mandracchia,[1,2] A. Finizio,[1] and P. Ferraro[1,*]

[1]CNR – ISASI, Institute of Applied Sciences & Intelligent Systems, Via Campi Flegrei 34, 80078 Pozzuoli (Naples), Italy
[2]Dipartment of Chemical Engineering, Materials, and Industrial Production, Univ. "Federico II", P.le Tecchio 80, 80100, Naples, Italy.

*Corresponding author: pietro.ferraro@cnr.it

As the world's population ages, cataract-induced visual dysfunction and blindness is on the increase. This is a significant global problem. The most common symptoms of cataracts are glared and blurred vision. Usually, people with cataract have trouble seeing or reading at distance or in low light and also their color perception is altered. Furthermore, cataract is a sneaky disease as it is usually a very slow but progressive process, which creates adaptation so that patients find it difficult to recognize. Moreover, for the doctors it can be very difficult to explain and give comprehensive answers to the patients' symptoms. We built and tested an optic device that uses egg albumen to mimic the optical degradation of the crystalline related cataracts and that is able to visualize how the cataract impairs vision. At best of our knowledge, it is the first experimental system developed at this aim. This can be a valuable tool, which can be of help in education for students in medical sciences as well as to provide a method to illustrate the patients how their vision is affected by cataract progression process.

Cataract is the world's leading cause of blindness. According to World Health Organization (WHO), it is responsible for the 51% of the estimated 39 million cases of blindness occurring worldwide [1]. It is a gradual and inevitable process, should we be fortunate enough to live a long life [2].

Cataract is a disease associated with aging and with photo-oxidative denaturation (and cross-linking) of lens crystallins and other proteins. Both cataract and aging of lens cells are associated with declining proteolytic capacity and diminished antioxidant protection. Lens aging and in vivo photo-oxidative stress can cause opalescence ("cataract"), cross-linking of crystallins, and diminished proteolytic capacity [3].

The most common symptoms of cataracts are glared and blurred vision. Usually reading or seeing at distance become difficult, need of more light is sensed and color vision is also affected. So, how much does cataract affect vision? How does a subject with cataract experience this gradual sight impairment?

Indeed, in the case of cataract, it is extremely difficult to evaluate the subjective vision loss as it depends on many factors such as location and spatial distribution. Besides, cataract's slow but progressive process creates adaptation and, hence, the sufferers do not easily recognize it. Moreover, doctors can find difficult to explain and give comprehensive answers to the patients' symptoms, as they have not experienced such multifaceted vision loss themselves.

How to make patients aware of their progressive vision lost? It would be quite useful for patients' awareness to have an experimental demonstrator for illustrating and displaying to patients how their vision is affected by cataract progression process. Here we show an experimental technique, including a simple eye cell model, that mimics the optical degradation of the crystalline related cataracts and that is able to visualize how the cataract impairs vision. This can be a valuable tool, which can be of help in education for students in medical science as well as to provide a method to illustrate the patients the vision disease thus enforcing their awareness in affording it.

In general, the quality of the image on the retina is degraded because of scattering, diffraction at the pupil, defocus due to accommodation or ametropia, and aberrations of the eye [4–8]. With age, the crystalline lens undergoes changes that result in a degradation of the optical performance of the human eye.

Over the course of life, UV in sunlight, chemicals, trauma or radiation experienced by each individual start to denature the proteins in lens [9]. Denaturing means that the nature of the proteins change and gradually the lens loses its transparency thus starting to scatter light. This explains how cataracts form. In this case, scattering becomes an important contributor of image degradation and the crystalline lens progressively loses transparency and becomes opalescent [6–8,10,11].

Pathological studies of cataractous lenses have revealed that cataracts are composed of protein aggregates that precipitate in eye lens cells [12–14]. The prevalent proteins within the eye lens are the crystallins. Lens transparency is thought to be maintained by a liquid-like, short range order present in highly concentrated solutions of these proteins [15–17]. In mammals, there are three classes of crystallins, of which ⍺-crystallin is the most abundant.

Age related modifications of ꞵ-crystallin induce alterations in lens crystallin interactions, which can be responsible for the increase in light scattering in old and cataractous lenses [18–21]. Alterations include both changes in the secondary structure and in the state of aggregation [22], [23]. They are triggered by lens cells exposition to elevated temperatures or other stress factors such as UV light or Cosmic radiation (e.g. it has been shown an increase of the risk of nuclear cataract in airline pilots [24]), which disrupt the liquid-like molecular order and promote the formation of large scattering particles [25–27].

Fig. 1: Sketch of the experimental setup (top) and technical draw of the

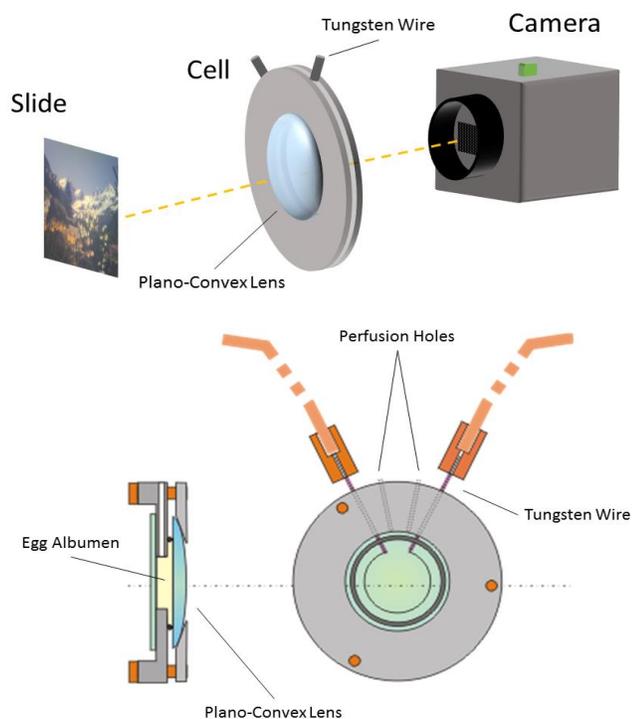

eye cell model (bottom).

Quite remarkably, temperature is a very simple way to reproduce many of the age-related lens changes. Simply exposing lenses to heat can replicate many changes that take place to the lens during our lifetimes, such as stiffness, loss of ꞵ-crystalline, membrane fluidity, binding of proteins to cell membranes and decomposition of membrane phospholipids. Time and temperature are linked variables. It has been shown that short-term exposure to 50°C in the laboratory can reasonably be used to study processes in an experimental period that in the normal eye may take decades at 35-42°C [28]. Given the rapidity with which major lens changes can be induced under laboratory conditions at 50°C, it would not be surprising if such variation in ocular temperature could indeed modulate the rate of change of protein denaturation in the lens.

This same denaturation process occurs to albumen when an egg is boiled. In the past, albumen have been used as phantom for the study of tissue response and photocoagulation induced by laser light [29,30]. In fact, it has a well-known behavior of protein denaturation as function of temperature and time, which is related to its scattering and adsorbing properties [31–33].

Here, we use egg albumen as realistic phantom material in order to reproduce the changes of visual impairment due to aging in very good analogy with the protein denaturation that occurs in the eye lens. We built a cell formed by a sealed box with transparent walls, one being a plano-convex lens and the other a glass microscope slide. The cell is filled with albumen through two small perfusion holes placed on top. On the inside, furthermore, a thin tungsten wire connected to a voltage generator provides heating by Joule effect. A LED lamp is used to illuminate a slide and form the image. To obtain homogenous illumination, a diffuser was placed between the slide and the lamp. Finally, a camera is placed in correspondence of the image plane of the eye cell (see Fig. 1).

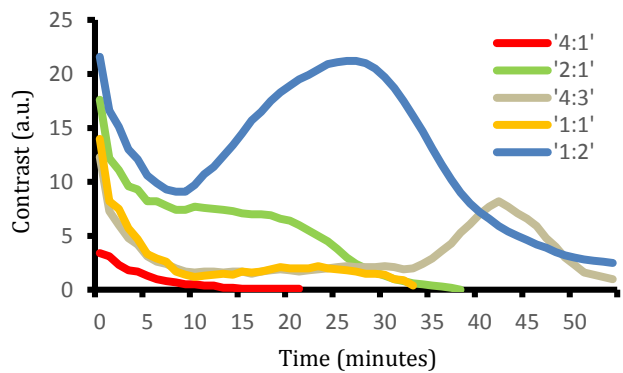

Fig. 2: Contrast behavior as function of time for different dilutions: 4:1 (red), 2:1 (green), 4:3 (gray), 1:1 (yellow), 1:2 (blue).

In order to reproduce the progressive effect of protein accumulation at the center of the crystalline in aging subjects, some arrangements must be made. In fact, to obtain sufficiently sharp images the albumen should be diluted to reduce the light scattering at room temperature. On the other hand, if it is diluted too much the effect of protein denaturation could become negligible. We have tested different albumen/water dilutions: 4:1, 2:1, 4:3, 1:1, and 1:2. To quantify the effect of scattering due to protein denaturation on the image quality, we used a test slide with a line pattern on it (2.5 lin/mm). For each dilution, we evaluated the image contrast as the standard deviation of the intensity, calculated on the area correspondent to the pattern (see Fig. 2). The working voltage was fixed at 1V.

Taking into account the time-behavior of contrast and image sharpness at initial stage of the measurements, the plot in Figure 2 shows a better compromise for dilution 1:1. Using these value of dilution, we simulated the cataract-like degradation of image quality using a real-life scenario (Figures 3, Visualization 1). As time passes, proteins start denaturing and accumulating in correspondence of the center of the eye cell. In figure 3c, the image shows at first a slight increase of contrast (+7%) with a concurrent loss of brightness (-4%). The image quality, then, keeps on worsening until the amount of denatured proteins fills the center of the cell and scattering is so strong that light does not reach the camera anymore.

Cataract induces also an alteration of color perception. The images formed by the sufferers eyes result dimmer and yellowed. This is due to scattering of light inside the crystalline, which is more severe for smaller wavelengths. This deviates blue light more the red one, producing, thus, a notable image yellowing. We have quantified color balance comparing the intensities of the three color channels of the camera, see Fig.3 insets. In Fig. 3b, the three channels are equally present in the image, although a slight prevalence of the blue one due to the sky. As time passes and scatter increases, the image loses definition and color balance starts to tend toward the red even though the effect is barely perceivable, see Fig. 3c. As the denaturation process goes on, albumen proteins scatter light to the extent that image formation is almost hindered and color balance is definitely uneven, Fig. 3d. This is consistent with the aforementioned cataract-related vision impairment symptoms such as dim vision and color yellowing.

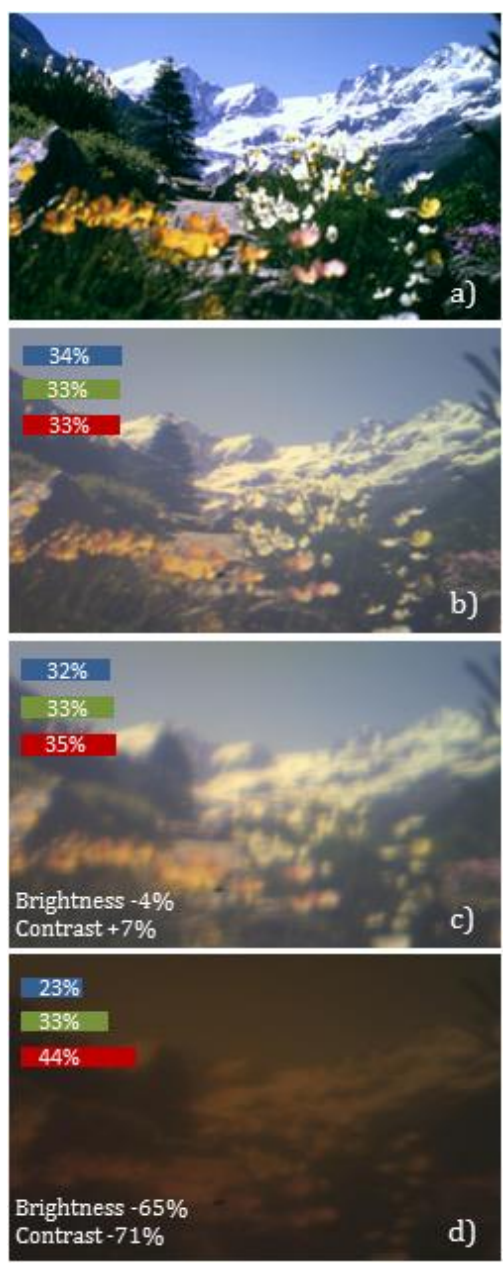

Fig. 3: Cataract-like image degradation of a photograph slide. a) Image of the slide taken using only water. Images of the slide taken using a 1:1 albumen/water dilution and applying a 1V tension. b)T=0min, c) T=2min, d) T=16min. The insets show the color distribution in the images.

This experimental setup, then, proved itself capable to reproduce the image degradation process due to denatured proteins accumulation, either for optical resolution, glare effects and color distortion. Given the right experimental parameters, i.e. 1V tension, diffused illumination, and albumen/water dilution within 2:1 and 1:1, it shows a gradual loss of image contrast due to light scattering. We have shown that it features all the main characteristics of sight weakening experienced by cataract sufferers: glaring, blurring, dim vision, color alteration, and, finally, blindness. The gradual and sequential appearance of such features seems to mimic well the steady and relentless effect of aging upon vision, making it suitable as a tool for the assessment and illustration of age-related evolution in vision impairment caused by cataracts.

**Supplementary material.** See Supplementary 1 for supporting content.